\documentclass[sigconf,screen]{acmart}
\setcopyright{rightsretained}
\acmPrice{}
\acmDOI{10.1145/3597926.3598059}
\acmYear{2023}
\copyrightyear{2023}
\acmSubmissionID{issta23main-p90-p}
\acmISBN{979-8-4007-0221-1/23/07}
\acmConference[ISSTA '23]{Proceedings of the 32nd ACM SIGSOFT International Symposium on Software Testing and Analysis}{July 17--21, 2023}{Seattle, WA, USA}
\acmBooktitle{Proceedings of the 32nd ACM SIGSOFT International Symposium on Software Testing and Analysis (ISSTA '23), July 17--21, 2023, Seattle, WA, USA}
\received{2023-02-16}
\received[accepted]{2023-05-03}

\AtBeginDocument{%
  \providecommand\BibTeX{{%
    \normalfont B\kern-0.5em{\scshape i\kern-0.25em b}\kern-0.8em\TeX}}}

\usepackage{booktabs}   
\usepackage{subcaption} 

\usepackage[utf8]{inputenc}
\usepackage[T1]{fontenc}
\usepackage[scaled=0.78]{beramono}
\usepackage{amsmath}
\usepackage[ruled,vlined,linesnumbered]{algorithm2e}

\usepackage{mathtools}
\usepackage{stmaryrd}
\usepackage{wasysym}
\usepackage{color}
\usepackage{xcolor,colortbl}
\usepackage{url}
\usepackage{listings}
\usepackage{paralist}
\usepackage{caption}
\usepackage{wrapfig}
\usepackage{enumitem}
\usepackage{multicol}
\usepackage{multirow}
\usepackage{makecell}
\usepackage{flushend}
\usepackage{bcprules}
\usepackage{textcomp}
\usepackage{tikz}
\usetikzlibrary{positioning,fit,calc,arrows.meta,arrows,decorations}
\usepackage{pdfpages}
\usepackage{cleveref}
\usetikzlibrary{matrix}
\usepackage{xspace}
\definecolor{light-gray}{gray}{0.85}
\usepackage{stackengine}
\usepackage{mathpartir}
\usepackage{graphicx}

\definecolor{ckeyword}{HTML}{000000}
\definecolor{ccomment}{HTML}{3F7F5F}
\definecolor{cstring}{HTML}{2A0099}

\lstdefinelanguage{Scala}%
{morekeywords={
  abstract, sealed, lazy,
  case,catch,char,class,%
  def,do,else,extends,final,finally,for,%
  if,import,implicit,%
  match,module,%
  new,null,undefined,%
  array,
  override,%
  package,private,protected,public,%
  for,public,return,super,%
  this,throw,trait,try,type,%
  val,var,%
  with,while,%
  object,
  let,skip,assert,then,fst,snd,idx,sum,prod,exists,forall,%
  yield,%
  define, null?, car, cdr
  },%
  sensitive,%
  moredelim=*[il][\bfseries]{\#\#\ },
  morecomment=[l]//,%
  morecomment=[s]{/*}{*/},%
  morestring=[b]",%
  showstringspaces=false%
}[keywords,comments,strings]%

\lstdefinelanguage{Effect}%
{morekeywords={
    effect, yield, return
  },%
  sensitive,%
  moredelim=*[il][\bfseries]{\#\#\ },
  morecomment=[l]//,%
  morecomment=[s]{/*}{*/},%
  morestring=[b]",%
  showstringspaces=false%
}[keywords,comments,strings]%

\lstdefinelanguage{LLVM}%
{morekeywords={
    define, i32, br, icmp, sub, call, mul, phi, ret, label
  },%
  sensitive,%
  moredelim=*[il][\bfseries]{\#\#\ },
  morecomment=[l];,%
  morecomment=[s]{/*}{*/},%
  morestring=[b]",%
  showstringspaces=false%
}[keywords,comments,strings]%

\lstdefinelanguage{CPP}%
{morekeywords={
    using, void, function, if, else, Cont, List
  },%
  sensitive,%
  moredelim=*[il][\bfseries]{\#\#\ },
  morecomment=[l]//,%
  morecomment=[s]{/*}{*/},%
  morestring=[b]",%
  showstringspaces=false%
}[keywords,comments,strings]%

\lstdefinestyle{small}{
  language=Scala,%
  mathescape=true,%
  aboveskip=2pt,
  belowskip=1pt,
  lineskip=-1pt,
  basewidth={0.6em, 0.45em},%
  basicstyle=\fontsize{7}{9}\selectfont\ttfamily,
  keywordstyle=\keywordstyle,
  commentstyle=\commentstyle,
  stringstyle=\stringstyle,
  literate={-->}{{$\to$}}3
           {->}{{$\mapsto$}}3
           {=>}{{$\Rightarrow ~$}}2
           {|-}{{$\ts$}}2
           {idx}{{$\#$}}1
           {sum}{{$\Sigma$}}1
           {array(}{{$\langle.\rangle$(}}3
           {σ}{{$\sigma$}}1
           {ρ}{{$\rho$}}1
           {→}{{$\to$}}1
           {λ}{{$\lambda$}}1
           {α}{{$\alpha$}}1
           {⊔}{{$\sqcup$}}1
           {⊓}{{$\sqcap$}}1
           {⊑}{{$\sqsubseteq$}}1
           {⊤}{{$\top$}}1
           {⊥}{{$\bot$}}1
           {×}{{$\times$}}1
           {τ}{{$\tau$}}1
           {ψ}{{$\psi$}}1
}

\lstdefinestyle{extrasmall}{
  language=Scala,%
  mathescape=true,%
  aboveskip=2pt,
  belowskip=1pt,
  lineskip=-1pt,
  basewidth={0.6em, 0.45em},%
  basicstyle=\fontsize{6}{8}\selectfont\ttfamily,
  keywordstyle=\keywordstyle,
  commentstyle=\commentstyle,
  stringstyle=\stringstyle,
  literate={-->}{{$\to$}}3
           {->}{{$\mapsto$}}3
           {=>}{{$\Rightarrow ~$}}2
           {|-}{{$\ts$}}2
           {idx}{{$\#$}}1
           {sum}{{$\Sigma$}}1
           {array(}{{$\langle.\rangle$(}}3
           {σ}{{$\sigma$}}1
           {ρ}{{$\rho$}}1
           {→}{{$\to$}}1
           {λ}{{$\lambda$}}1
           {α}{{$\alpha$}}1
           {⊔}{{$\sqcup$}}1
           {⊓}{{$\sqcap$}}1
           {⊑}{{$\sqsubseteq$}}1
           {⊤}{{$\top$}}1
           {⊥}{{$\bot$}}1
           {×}{{$\times$}}1
}

\definecolor{listingbg}{RGB}{240, 240, 240}

\newcommand{\commentstyle}[1]{\color{ccomment}\itshape{#1}}
\newcommand{\keywordstyle}[1]{\color{ckeyword}\bfseries{#1}}
\newcommand{\stringstyle}[1]{\color{cstring}\text{#1}}

\lstnewenvironment{listing}{\lstset{language=Scala}}{}
\lstnewenvironment{listingtiny}{\lstset{language=Scala,basicstyle=\scriptsize\ttfamily}}{}

\newcommand{\code}[1]{\lstinline[language=Scala,columns=fixed,basicstyle=\ttfamily]|#1|}

\definecolor{clcolor}{rgb}{0.5,0.7,0.9}
\definecolor{kscolor}{rgb}{0.9,0.1,0.1}
\definecolor{rbcolor}{rgb}{0.7,0.4,0.7}
\definecolor{nkcolor}{rgb}{0.4,0.7,0.7}
\definecolor{rfcolor}{rgb}{0.56, 0.0, 1.0}
\definecolor{cscolor}{rgb}{0.1, 0.4, 1.0}
\definecolor{jpcolor}{rgb}{0.36, 0.54, 0.66}

\usepackage{listings, xcolor}

\definecolor{verylightgray}{rgb}{.97,.97,.97}

\lstdefinelanguage{Solidity}{
	keywords=[1]{anonymous, assembly, assert, balance, break, call, callcode, case, catch, class, constant, continue, constructor, contract, debugger, default, delegatecall, delete, do, else, emit, event, experimental, export, external, false, finally, for, function, gas, if, implements, import, in, indexed, instanceof, interface, internal, is, length, library, log0, log1, log2, log3, log4, memory, modifier, new, payable, pragma, private, protected, public, pure, push, require, return, returns, revert, selfdestruct, send, solidity, storage, struct, suicide, super, switch, then, this, throw, transfer, true, try, typeof, using, value, view, while, with, addmod, ecrecover, keccak256, mulmod, ripemd160, sha256, sha3}, 
	keywordstyle=[1]\color{blue}\bfseries,
	keywords=[2]{address, bool, byte, bytes, bytes1, bytes2, bytes3, bytes4, bytes5, bytes6, bytes7, bytes8, bytes9, bytes10, bytes11, bytes12, bytes13, bytes14, bytes15, bytes16, bytes17, bytes18, bytes19, bytes20, bytes21, bytes22, bytes23, bytes24, bytes25, bytes26, bytes27, bytes28, bytes29, bytes30, bytes31, bytes32, enum, int, int8, int16, int24, int32, int40, int48, int56, int64, int72, int80, int88, int96, int104, int112, int120, int128, int136, int144, int152, int160, int168, int176, int184, int192, int200, int208, int216, int224, int232, int240, int248, int256, mapping, string, uint, uint8, uint16, uint24, uint32, uint40, uint48, uint56, uint64, uint72, uint80, uint88, uint96, uint104, uint112, uint120, uint128, uint136, uint144, uint152, uint160, uint168, uint176, uint184, uint192, uint200, uint208, uint216, uint224, uint232, uint240, uint248, uint256, var, void, ether, finney, szabo, wei, days, hours, minutes, seconds, weeks, years},	
	keywordstyle=[2]\color{teal}\bfseries,
	keywords=[3]{block, blockhash, coinbase, difficulty, gaslimit, number, timestamp, msg, data, gas, sender, sig, value, now, tx, gasprice, origin},	
	keywordstyle=[3]\color{violet}\bfseries,
	identifierstyle=\color{black},
	sensitive=false,
	comment=[l]{//},
	morecomment=[s]{/*}{*/},
	commentstyle=\color{gray}\ttfamily,
	stringstyle=\color{red}\ttfamily,
	morestring=[b]',
	morestring=[b]"
}

\newcommand{\tool}{\textsc{ItyFuzz}\xspace}

\newcommand{\cutreentrancy}{\iffalse}

\begin{document}

\title{\tool: Snapshot-Based Fuzzer for Smart Contract}

\author{Chaofan Shou}
\email{shou@berkeley.edu}
\affiliation{%
  \institution{UC Berkeley}
  \city{Berkeley}
  \state{CA}
  \country{USA}
  \postcode{94709}
}

\author{Shangyin Tan}
\email{shangyin@berkeley.edu}
\affiliation{%
  \institution{UC Berkeley}
  \city{Berkeley}
  \state{CA}
  \country{USA}
  \postcode{94709}
}
\author{Koushik Sen}
\email{ksen@berkeley.edu}
\affiliation{%
  \institution{UC Berkeley}
  \city{Berkeley}
  \state{CA}
  \country{USA}
  \postcode{94709}
}

\lstMakeShortInline[keywordstyle=,%
                    flexiblecolumns=false,%
                    mathescape=false,%
                    basicstyle=\ttfamily]@



\begin{abstract}
Smart contracts are critical financial instruments, and their security is of utmost importance. However, smart contract programs are difficult to fuzz due to the persistent blockchain state behind all transactions. Mutating sequences of transactions are complex and often lead to a suboptimal exploration for both input and program spaces. In this paper, we introduce a novel snapshot-based fuzzer \tool for testing smart contracts. In \tool, instead of storing sequences of transactions and mutating from them, we snapshot states and singleton transactions. To explore interesting states, \tool introduces a dataflow waypoint mechanism to identify states with more potential momentum. \tool also incorporates comparison waypoints to prune the space of states. By maintaining snapshots of the states, \tool can synthesize concrete exploits like reentrancy attacks quickly. Because \tool has second-level response time to test a smart contract, it can be used for on-chain testing, which has many benefits compared to local development testing. Finally, we evaluate \tool on real-world smart contracts and some hacked on-chain DeFi projects. \tool outperforms existing fuzzers in terms of instructional coverage and can find and generate realistic exploits for on-chain projects quickly.

\end{abstract}




\maketitle

\section{Introduction}

Smart contract auditing has become a billion-dollar industry with the increasing adoption of Web 3.0 technology and the growing number of attacks. Smart contracts are programs deployed on the blockchain and can accept transactions from any party. Transactions are calls to public functions or token transfers to the deployed smart contract. Each transaction serves as an input and can modify the smart contract's state. Auditing means ensuring a smart contract has no vulnerabilities to losing assets stored inside it. Smart contracts can be audited by automated fuzzing tools, but all fuzzing tools for smart contracts only support testing in the local deployment, not on the blockchain. Fuzzing on the blockchain requires high exploration speed for a given state, because (1) the state of the blockchain is constantly changing, and (2) attacker exploits can happen at any time. Current smart contract fuzzers are not efficient enough to fetch the states from the blockchain and finish the auditing quickly. However, on-chain auditing, or directly performing fuzz testing with the states fetched from the blockchain continuously, is critical in the following scenarios. First, when certain code locations are only reachable from specific on-chain states, local or development setting fuzzing is useless. Second, modern smart contracts often leverage external contracts as sources of information. With on-chain auditing, we can pull these data from the blockchain dynamically in real-time.

Despite all the benefits, on-chain auditing becomes useless when the fuzzers can not provide a real-time response, as the ultimate goal is to pause the contract \textit{before} the attackers identify or conduct the attack. Existing fuzzers are mostly under-optimized for response time. For achieving full coverage of smart contracts, existing fuzzers have to spend a significant period (hours or days) while on-chain auditing needs second-level response time. In this paper, we propose a new \text{\it  snapshot-based fuzzing} technique and develop a fuzzer called \tool. \tool can achieve high coverage over code and states of smart contracts in just a few seconds and thus support on-chain auditing on many smart contracts.

Although the amount of source code of most smart contracts is trivial compared to complex software like browsers and operating systems, they are stateful and have complex dependencies with other smart contracts, which makes them hard to fuzz. Aiming to tackle the stateful nature of smart contracts, some previous works start from a fresh state for each fuzz run with a sequence of transactions as input. During the mutation phase, parts of this sequence are mutated. This way, existing tools have high overhead on re-executing transactions to return to a previous state. For exploring programs with a deeper state, which needs to be built up with several transactions, re-execution cost grows linearly. Additionally, existing tools only have feedback mechanisms for transactions but not for states, yet states and transactions have different exploration difficulties. We argue that the interestingness of states is equally important as  transactions for stateful fuzzing, and such feedback mechnisms to choose interesting states to explore is non-existent in current stateful fuzzing tools.


Instead of re-executing inputs to build up previous states, we propose snapshot-based fuzzing. Snapshots are essentially replicas of intermediate states built from some transactions. By storing all interesting snapshots into a state corpus, \tool can ``time travel'' to previous states with constant complexity ($O(1)$). Time traveling allows for efficient exploration for search space of both transactions and states. To support fast snapshotting, we refactor an existing Ethereum Virtual Machine (EVM) implementation. However, storing all snapshots into the corpus is still not practical due to limited runtime memory resources, while the number of snapshots increases linearly with the total execution time. The size of the stored snapshots could grow to several gigabytes in a few seconds. To resolve this issue and prioritize explorations of the most interesting states, we design two feedback mechanisms (i.e., waypoints) to classify interesting states and a corpus pruning technique to reduce the number of interesting states when necessary.

By reducing overhead from re-execution and improving the feedback mechanism, we can reduce fuzzer time significantly so that most vulnerabilities can be uncovered instantly. Thus, \tool can support the on-chain auditing goal, which requires fast response time to front run the attackers. 

\cutreentrancy

\paragraph{Concrete Exploit Synthesis for Reentrancy Attack}
Another critical path to support on-chain auditing is not only to detect but also to generate concrete exploits. Concrete exploits are essential for on-chain auditing because on-chain auditing requires instant action-making, usually through automation, to reduce the reaction time between the vulnerability appearing and the contract being paused. Any false positive results are unacceptable since pausing a smart contract is equivalent to a denial of service. Contract pausing should only occur when the contract has confirmed vulnerabilities and these vulnerabilities can be reproduced. 

All existing symbolic execution and fuzzing tools can generate concrete transactions for triggering simple vulnerabilities like integer overflow and control flow hijack, for which exploits are simple. However, none of them could synthesize an attack for a common yet dangerous vulnerability - the reentrancy vulnerability (i.e., leveraging external calls to arbitrary addresses to re-enter the contract when code is only partially executed, see \Cref{subsec:reentrancy}). Instead, they mark all external calls followed by state-changing operations as having reentrancy vulnerabilities. This over-simplification yields a significant number of false positives.

\tool efficiently resolves this problem via snapshotting. Since the reentrancy attack is to re-enter the contract, we can explore all potential re-entering locations via fuzzing. By snapshotting the partial state (i.e., incomplete execution) before an external call happens and using feedback mechanisms designed explicitly for states, \tool can efficiently build up complex reentrancy attack exploit. To further optimize for reentrancy exploit synthesis, we create a new feedback mechanism that tracks the dataflow of the current execution. 

\fi

Although \tool focuses on smart contract fuzzing, the snapshot fuzzing idea we developed can be applied to other domains as well. For example, modern hardware designs are highly stateful programs, where some bugs are only reachable through specific sequences of input signals. Traditional software fuzzing does not work well when the bug-triggering signal sequence becomes longer, but snapshot-based fuzzing can discover the more interesting state spaces that potentially lead to the desired bug. The only requirements for our current snapshot fuzzing algorithm are an efficient representation of the program state and the ability to observe the state during program execution. \\



\noindent\textit{Contributions.} In summary, we make the following contributions:
\begin{itemize}[leftmargin=1.5em]
\item We present a novel snapshot-based fuzzing algorithm to reduce re-execution
overhead for stateful smart contract fuzzing (\Cref{subsec:snapshot}).
\item We create new waypoint mechanisms optimized for prioritizing the exploration of interesting snapshot states, allowing efficient program exploration:
\begin{enumerate}
    \item Dataflow waypoint (\Cref{sec:df}) evaluates the interestingness of
    states based on ``future'' memory load.
    \item Comparison waypoint (\Cref{subsec:cmp}) compresses the state corpus by
    probabilistic sampling and hard comparison feedback.
\end{enumerate}
\item We develop a fast and efficient smart contract fuzzer \tool (\Cref{sec:impl}) and demonstrate its effectiveness (\Cref{sec:eval}).
\item Based on \tool, we propose a new auditing method for smart contracts to
conduct testing based on state fetched from the blockchain on the fly (\Cref{subsec:onchain}). Using
this method, we detect and reproduce exploits of on-chain projects worth millions of dollars.
\end{itemize}

\section{Background}

\subsection{Fuzzing}

\paragraph{Coverage-guided fuzzing}
Fuzz testing or fuzzing is a technique to automatically find vulnerabilities and bugs in
software by providing random inputs to the target program. To better explore the
target program, a fuzzer often employs heuristics and feedback from test executions to generate lots of new interesting inputs by mutating existing test inputs.  To this date, coverage-guided fuzzers have found numerous bugs in many real-world software systems.

We show a simplified algorithm for coverage-guided fuzzer in~\Cref{fig:cgf}. The
fuzzer starts with an initial corpus $\mathcal{I}$, which contains a set of
initial inputs, where in each iteration, an input $i$ from the corpus is
selected.  $i$ is then randomly mutated to produce several mutated inputs. The fuzzer then executes each mutated
input $i_m$ and checks whether the execution is interesting.  An execution is interesting if it covers new coverage points that have not been covered by existing inputs.  If the execution is
interesting, the fuzzer adds the mutated input $i_m$ to the corpus $\mathcal{I}$
and updates the coverage information.  The addition of $i_m$ to the corpus ensures that the input gets further chance to mutate in future.  Thus the fuzzer can explore the
target programs more efficiently using coverage feedback while ignoring redundant (or uninteresting) inputs. In \Cref{subsec:waypoint}, we describe a more general feedback called \emph{Waypoint} first proposed in~\cite{DBLP:journals/pacmpl/PadhyeLSSV19}.

\begin{figure}[ht]
   \begin{lstlisting}[language=Python, mathescape, numbers=left, xleftmargin=3.5em]
   $\mathcal{I}$ <- $\ initial\_corpus$
   coverage <- $\emptyset$
   
   while true:
      i <- random.choice($\mathcal{I}$)
      $\mathcal{I}$ <- $\ \mathcal{I}$ \ {i}
      for $i_m$ in Mutation(i):
         f <- execute($i_m$)
         if f increases coverage:
            $\mathcal{I}$ <- $\ \mathcal{I}\ \cup $ t
            coverage <- coverage $\cup$ f
     
   \end{lstlisting}
   \caption{Coverage-guided fuzzing algorithm (simplified)}\label{fig:cgf}
\end{figure}

\paragraph{Fuzz Smart Contract}
Although fuzzing techniques have been widely adopted to test traditional
software systems, smart contracts pose new challenges to the current
coverage-guided fuzzing techniques. As discussed in
\Cref{subsec:smart-contract}, smart contracts are stateful programs. Because
smart contracts have access to persistent memory on the blockchain, constructing
inputs individually to test smart contracts in a given persistent state is ineffective.
To address this issue, smart contract fuzzers must produce a sequence of
inputs (transactions) to test the smart contract.

\subsection{Waypoints}\label{subsec:waypoint}

Fuzzfactory \cite{DBLP:journals/pacmpl/PadhyeLSSV19} summarized a generalized feedback mechanism called waypoint. Waypoints are intermediate inputs that provide interesting feedback after executing the target program. For example, in coverage-guided fuzzing algorithm \Cref{fig:cgf}, new inputs (waypoints) are recorded if running the target program produces new coverage. However, waypoints are not limited to the coverage points---some other common waypoints include execution time, memory usage, and distance between two compared values waypoint. To implement customized waypoints, we need to collect other targeted, dynamic information during the execution of the program and provide a new predicate function \code{is\_interesting} to replace line 9 in \Cref{fig:cgf}. Unlike traditional fuzzers that use waypoints to test the interestingness of inputs, in \tool, we design novel state comparison and dataflow waypoints to decide if output states are interesting for snapshot-based fuzzing. We will discuss dataflow and comparison waypoints for states in more detail in \Cref{sec:df} and \Cref{subsec:cmp}. 

\subsection{Smart Contract} \label{subsec:smart-contract}

Smart contracts are computer programs deployed on the blockchain. The inputs to
smart contracts are typically called transactions. Once a transaction is
executed and posted on the blockchain, it is immutable and irreversible, and the
state change caused by the transaction adds to the persistent state of the blockchain. Because
blockchain and smart contracts are immutable and decentralized, many
applications, including decentralized finance, voting, gaming, etc., have been
built with smart contracts.

As we discussed before, smart contracts are on-chain computer programs. Many programming languages can be used to write smart contracts. Among them,
Solidity is arguably one of the most popular languages for writing smart
contracts. Solidity is a high-level language inspired by famous languages like
Javascript, Python, and C++. Solidity programs are deployed and executed on a
special blockchain called Ethereum. To run Solidity programs on the Ethereum
blockchain, one has to compile Solidity programs into Ethereum Virtual Machine (EVM)
bytecode. We show one example of a Solidity program in \Cref{fig:example}, and we will explain it in more detail in \Cref{sec:example}.

\cutreentrancy

\Cref{fig:reentrancy}.
The variable @balances@ is part of the persistent state of the smart contract.
The possible transactions can be calling @withdraw()@, and if precondition passes for this transaction, state change like @balances[msg.sender] = 0@ becomes persistent.

\begin{figure}[ht]
\begin{lstlisting}[language=Solidity, xleftmargin=3.5em, xrightmargin=0.5em]
contract UnsafeContract {
  mapping(address => uint) public balances;

  function withdraw() public {
     require(balances[msg.sender] > 0); // msg.sender is the address of the caller
     uint amount = balances[msg.sender];
     // sending value to sender and call a fallback (external) function with parameter ""
     (bool success, ) = msg.sender.call{value: amount}("");
     
     require(success);
     balances[msg.sender] = 0;
  }
}
\end{lstlisting}
\caption{A classical reentrancy bug example in Solidity. Line 8 transfers the balance and calls an external function, which may enter the \code{widthraw()} function again, causing a reentrancy bug}\label{fig:reentrancy}
\end{figure}

\subsection{Reentrancy Attack} \label{subsec:reentrancy}
One common, critical, and hard-to-find vulnerability in a smart contract is the reentrancy bug. A reentrancy attack happens when a smart contract calls an external contract, and the external contract calls back into the original contract. For example, in \Cref{fig:reentrancy}, the contracts hold the balance of tokens of the users, and the @withdraw@ function allows the user (@msg.sender@) to withdraw all their balance. However, since at line 8, the contract transfers the balance to @msg.sender@ and calls an external function, which yields the control, the malicious external contract can call the @withdraw@ function again to withdraw the balance multiple times since @balances[msg.sender]@ is never set to 0. Note since @withdraw@ sets the balance after calling the external contract, all reentrance calls to @withdraw@ will succeed, and the @UnsafeContract@ will be drained.

This type of reentrancy bug is notoriously hard to detect and exploit because real-world reentrancy exploits often contain many consecutive transactions. In \Cref{subsec:reentrancy_syn}, we will discuss how our snapshot-based fuzzing technique can help synthesize reentrancy exploits in smart contracts.
\fi

\section{Motivating Example}\label{sec:example}

To see why current sequence-based fuzzers fail to exploit some vulnerabilities, consider a simple yet realistic smart contract program in \Cref{fig:example} with s a single state variable @counter@. The function @incr@ increases the @counter@ by one when the argument is smaller than the current @counter@, and similarly, @decr@ decreases the @counter@ when the argument is greater than @counter@. Function @buggy@ introduces a bug when @counter@ reaches some constant @T@.

The inputs to this program are sequences of transactions, where each transaction is simply the function to execute and its corresponding parameters. For example, when @T == 2@, a bug-triggering transaction sequence is @[(incr, 0), (incr, 1), (buggy, )]@. This sequence first calls @incr@ with valid inputs two times to increase the @counter@ to @2@. Then, the last transaction calls @buggy@, which will trigger the bug.

Although the above input sequence seems simple and easy to synthesize, constructing a more complex transaction sequence when @T@ is large is non-trivial. The bug-triggering transaction sequence can be very long and complex in real-world scenarios. For instance, hackers leveraged six transactions, each with on average 40 function calls to build up the state that makes Team Finance decentralized finance (DeFi) platform vulnerable to attack \cite{TeamFinance}.

\begin{figure}
\begin{lstlisting}[language=Solidity, xleftmargin=3.5em, xrightmargin=0.5em]
contract SimpleState {
    int256 counter = 0;

    function incr(int256 x) public {
        require(x <= counter);
        counter += 1;
    }

    function decr(int256 x) public {
        require(x >= counter);
        counter -= 1;
    }

    function buggy() public {
        if (counter == T) {
            bug!();
        }
    }
}
    \end{lstlisting}
    \caption{Smart contract with a simple persistent state \code{counter}}
    \label{fig:example}
\end{figure}

\begin{figure}
        \centering
        \includegraphics[width=230pt]{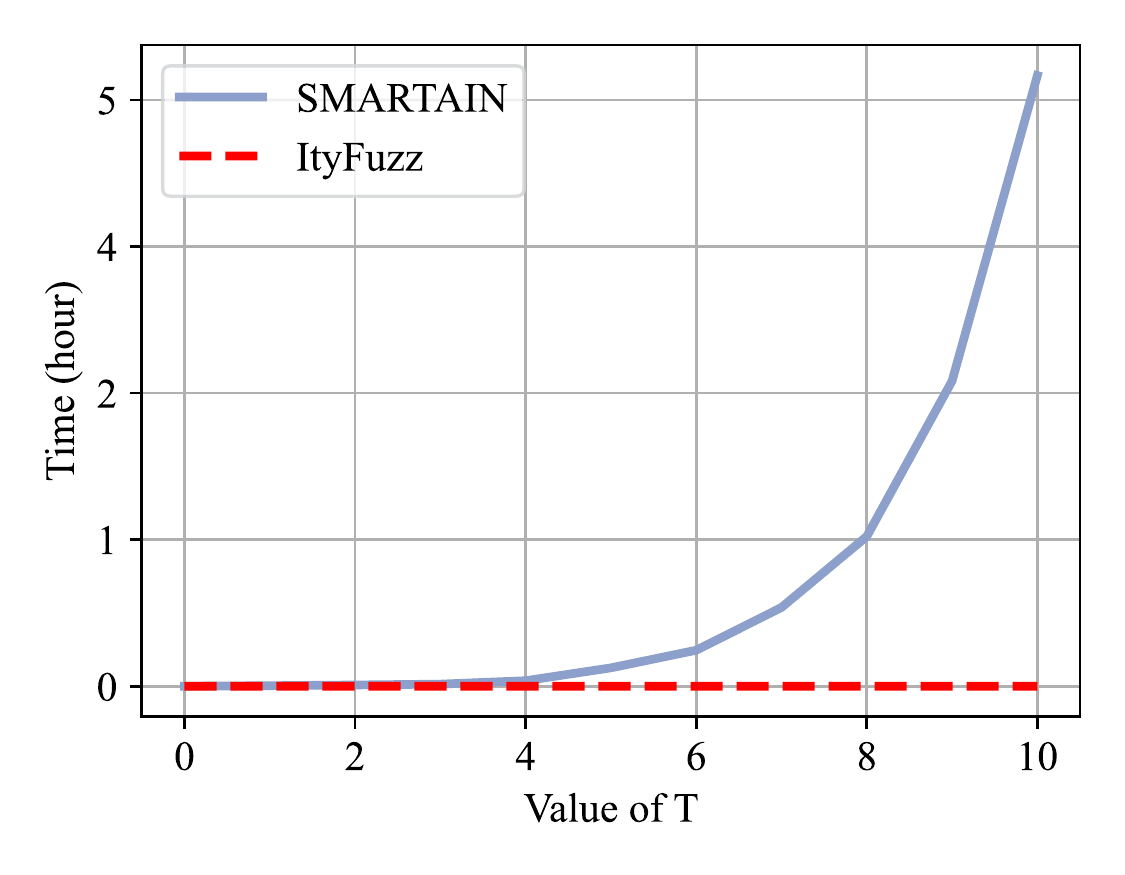}
        \caption{Time (hours) to find a bug-triggering transaction sequence for \code{SimpleState} with different \code{T} values}
        \label{fig:executecontract}
   
\end{figure}

\begin{figure}
    
        \centering
        \includegraphics[width=240pt]{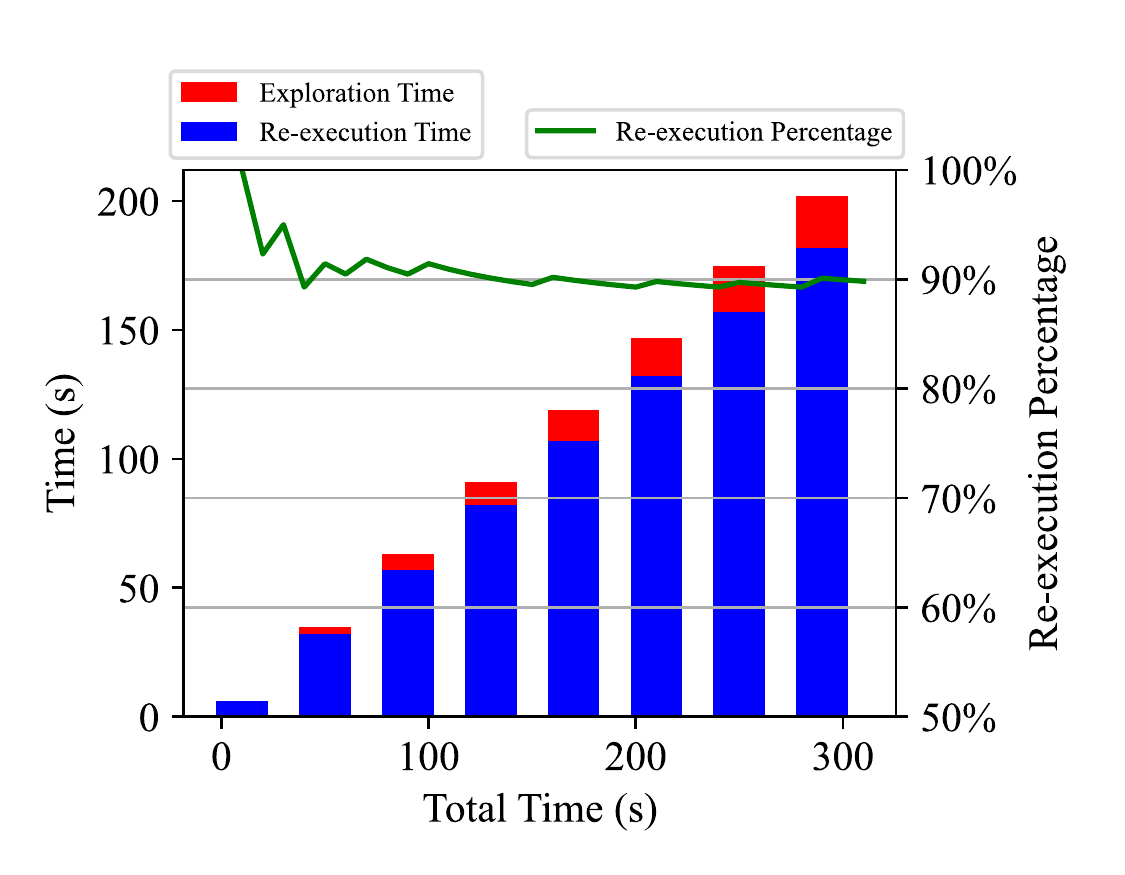}
        \caption{Re-execution time (y-axis) and percentages when running  \code{SimpleState} on SMARTIAN}
        \label{fig:reexecution}
\end{figure}



When exploring the transaction sequence space of this simple smart contract, previous state-of-the-art smart contract fuzzers like SMARTIAN \cite{DBLP:conf/kbse/0001K0GGC21} fail to quickly detect a bug as @T@ increases (\Cref{fig:executecontract}). This is because when evaluating a transaction sequence, the fuzzer needs to re-execute the entire sequence of transactions from the very beginning, including deploying the transaction to be tested. Time spent executing a transaction in an arbitrary state needs to include the time spent re-executing the sequence of transactions required to reach the arbitrary state. 

To empirically study how re-execution slows down the overall performance, we run SMARTIAN with default mode on the @SimpleState@ contract and set @T = 10@. We recorded the re-execution time and their percentage to the total fuzzing time in \Cref{fig:reexecution}. Re-execution times are recorded when the EVM executor executes a seen @(state, transaction)@ pair, and exploration times are recorded otherwise. In SMARTIAN, re-execution takes more than 90 percent of the total fuzzing time.

The amortized re-execution time grows exponentially as the length of the sequence of transactions grows linearly. Fortunately, the re-execution time can be eliminated if the state reached after executing a sequence of transactions can be memoized. However, memoization requires saving the number of states that is exponential in the number of transaction sequences. The key insight in \tool is that we can only memoize a set of ``interesting states'', called \emph{snapshots}, instead of memoizing all intermediate states and using these interesting states only to explore new states without re-execution. The ``interestingness'' of a state is defined using two novel waypoint ideas. We discuss the snapshot-based fuzzing technique in detail in \Cref{subsec:snapshot}.


\section{Methodology}
This section introduces our methodology to build \tool. We first describe the overall architecture of \tool (\Cref{subsec:arch}). Then we describe three crucial building blocks for \tool: snapshot-based fuzzing (\Cref{subsec:snapshot}), dataflow waypoint (\Cref{sec:df}), and comparison waypoint (\Cref{subsec:cmp}). 

\begin{figure*}[ht]
      \centering
      \includegraphics[width=.60\textwidth, trim={3cm, 2.5cm, 1.5cm, 2cm}, clip]{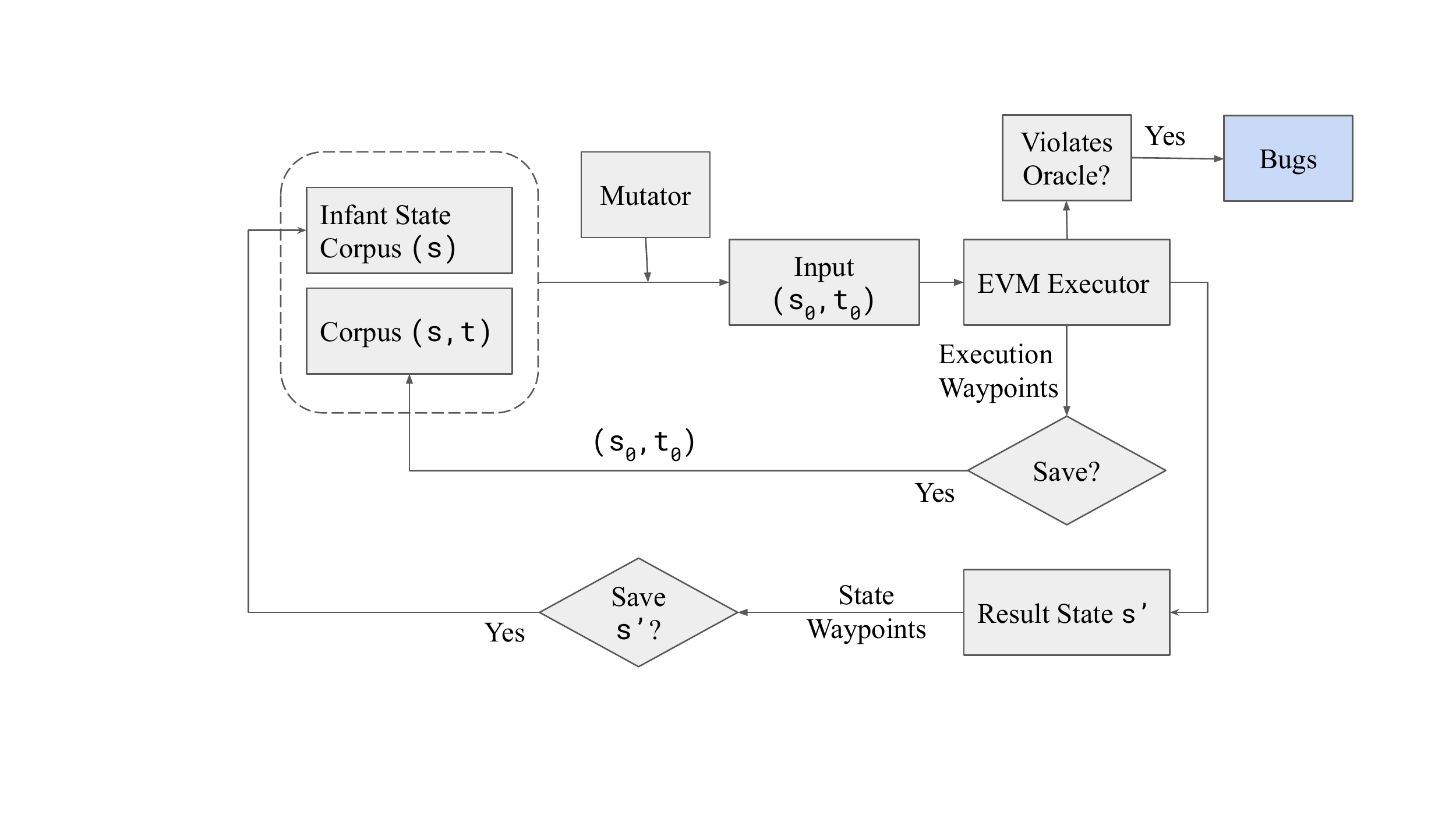}
      \caption{Architecture of \tool}
      \label{fig:arch}
\end{figure*}

\subsection{\tool Architecture}\label{subsec:arch}
We present the general architecture of \tool in \Cref{fig:arch}. To understand the architecture of \tool, recall that Solidity smart contract programs are executed on Ethereum Virtual Machine (EVM). EVM can be viewed as a function ${\rm\it EVM}: (S \times T) \to S$, which maps a state $s \in S$ and a transaction $t \in T$ to a new state in $S$ which we denote by $s_t$.  Note that the target program is an immutable part of the initial state $s$. 
Like conventional coverage-guided fuzzers, \tool starts the fuzzing loop with a corpus of seed inputs (shown in the left-top corner of \Cref{fig:arch}). In each iteration, a transaction and state pair is selected from the corpus and mutated to generate a pair, say (@s@$_0$, @t@$_0$). The mutated input is executed by the EVM executor. After the execution, \tool utilizes the execution waypoints to decide if the trace collected during the execution is ``interesting".  An input is said to be interesting if it increases the coverage of some waypoint. If yes, \tool adds the mutated input pair (@s@$_0$, @t@$_0$) to the corpus. Unlike conventional converage-guided fuzzers, the EVM executor also returns the updated state. We employ the state waypoints to save \textit{snapshots} of interesting result states.

\subsection{Snapshot-Based Fuzzing}\label{subsec:snapshot}
In smart contract executions, a state $s$ can be constructed by executing a sequence of transactions from an initial state.  To travel back to some intermediate state, a common practice for existing stateful fuzzers is to re-execute the transactions to construct the specific state from the initial state. Instead of re-executing the previous transactions, we directly snapshot the state and save the unique states. We maintain a corpus $\mathcal{C}$ that stores pairs of the form $(s, t)$, where $s \in S$ and $t \in T$. Specifically, given an execution ${\rm\it EVM}(s, t) \mapsto s'$, where $s' \in S$ is the new state after execution of $t$ on $s$, the pair ($s$, $t$) is added to the corpus when the combination of feedback (waypoints) for $\mathcal{C}$ reported by the execution of the transaction on that state is interesting. We employ multiple waypoint mechanisms, including coverage-guided feedback, dataflow waypoint (\Cref{sec:df}), and comparison waypoint (\Cref{subsec:cmp}). Since \tool always adds interesting pairs to the corpus, future exploration with the mutants of these pairs may also be rewarding.

Since \tool adds the state before a transaction to the corpus $\mathcal{C}$, the state obtained after executing the transaction is lost.  Therefore, if \tool keeps mutating the states from the corpus, it is only going to explore the initial state with random transactions. The subsequent states built with transactions on the initial states are never explored, undermining the purpose of stateful fuzzing, which is to explore all potential states. To make our exploration more efficient and to avoid re-execution, \tool maintains a separate corpus of states to memorize the states after execution, which we call the ``infant state corpus'', say $\mathcal{C}_{s}$. Specifically, given an execution ${\rm\it EVM}(s, t) \mapsto s'$, the state $s'$ is added to the infant state corpus when \tool finds it interesting using a combination of waypoints. Note that the waypoints for the infant state corpus are different from that of $\mathcal{C}$---they determine whether the current execution is interesting, but waypoints for the infant state corpus determine whether such a state can lead to future interesting executions.  We discuss how to compute the waypoints for the infant state-corpus in \Cref{sec:df} and \Cref{subsec:cmp}.

\begin{figure}[t]
  \begin{lstlisting}[language=Python, mathescape, numbers=left, xleftmargin=3.5em]
  $\mathcal{C}$ <-$\ $Corpus() $\ $ # transaction and state pair corpus
  $\mathcal{C}_s$ <-$\ $Corpus() $\ $ # infant state corpus

  while true:
     ($s$,$t$) <-$\ $Next($\mathcal{C}$)
     if Random(0, 1) > P:
        $t_{\rm \it mut}$ <-$\ $Mutation($t$) # mutate transaction
        $s_{\rm \it mut}$ <-$\ $s
     else 
        $s_{\rm \it mut}$ <-$\ $Next($\mathcal{C}_{s}$) # mutate state by fetching from infant state corpus
        $t_{\rm \it mut}$ <-$\ $t
     
     # execute transaction $t_{\rm\it mut}$ on state $s_{\rm \it mut}$
     # $f$ represents waypoints, $s'$ is the resulting state
     $f,\ s'$ <- ${\rm\it EVM}$($s_{\rm \it mut}$, $t_{\rm \it mut}$) 
     if $W_{(s,t)}$(f) is interesting: # get feedback for waypoints of execution
         $\mathcal{C}$ <- $\ \mathcal{C}\ \cup $ ($s_{\rm \it mut}$, $t_{\rm \it mut}$)
     if $W_{(s)}$(f, $s'$) is interesting: # get feedback for waypoints of infant state corpus
         $\mathcal{C}_{s}$ <- $\ \mathcal{C}_{s}\ \cup $ $s'$
  \end{lstlisting}
  \caption{\tool Fuzzing Algorithm}\label{fig:ouralgo}
  \end{figure}

The pseudo-code in \Cref{fig:ouralgo} summarizes \tool{}'s snapshot-based fuzzing algorithm .  During each iteration of the fuzz loop, the fuzzer first picks a state and transaction pair from the corpus $\mathcal{C}$. The fuzzer then either mutates the transaction $t$ structurally to $t_{\rm \it mut}$ or replaces the state $s$ with a state, say $s_{\rm \it mut}$, stored in the infant state corpus $\mathcal{C}_s$. The new mutant is always sound (i.e. reachable in the input space) because the selected state from the infant state corpus can be constructed by a sequence of transactions, and the state allows the execution of a new transaction. The mutant is executed by the EVM, which yields a new state. After the execution, \tool saves the state and transaction pair $(s_{\rm \it mut}, t_{\rm \it mut})$ to corpus $\mathcal{C}$ if the observed waypoints are interesting. \tool adds the the resulting state $s'$ to the infant state corpus $\mathcal{C}_{s}$ if $s'$ may lead to a potentially interesting executions with new transactions.  In \Cref{sec:df} and \Cref{subsec:cmp}, we will discuss the details of two types of state waypoints, dataflow waypoint and comparison waypoint, to decide if $s'$ are interesting.

\subsection{Dataflow Waypoint}
\label{sec:df}

Defining waypoints for measuring the interestingness of an input during execution has been explored by Padhye et al. in FuzzFactory \cite{DBLP:journals/pacmpl/PadhyeLSSV19} (e.g., recording coverage and memory-related instruction call patterns). Yet, there has been no work on how to define waypoints for states. The goal for the waypoints is to save resulting states in the corpus only if future executions from these states are interesting. Therefore, testing if a state is interesting also requires semantic information about the state in order to capture "future interestingness".

We define two waypoints for states. One is called \emph{dataflow waypoint}. With dynamic dataflow information, we know a memory location is interesting if it appears as the argument of the @load@ instruction in the future. If a state change contains unique writes to interesting memory locations, we can conclude that the state is interesting for future exploration. We leverage bytecode instrumentation to conduct dynamic dataflow analysis by observing @load@ and @store@ instructions on the fly. Note that conventional way to perform dataflow analysis is through static analysis of the source code. However, static analysis tools do not work well with smart contract fuzzers because smart contract may call external contracts dynamically. However, the static dataflow information gained only from the target contract is not sufficient.




\SetKwComment{Comment}{/* }{ */}

\begin{algorithm}
\KwResult{$L$, $S$}
\caption{Algorithm for dataflow waypoint instrumentation}\label{alg:df1}
\While{EVM is Executing}{
  \If{Current Instruction is load(Loc)}{
    $L$ (Loc \% $\text{MAP\_SIZE}$)  $\gets$ true \;
  }{\If{Current Instruction is store(Loc, Value)}{
      $S$(Loc \% $\text{MAP\_SIZE}$)(BucketOf(Value)) $\gets$ true\;
    }
  }
}
\end{algorithm}

\begin{algorithm}
\caption{Algorithm for dataflow waypoint evaluation}\label{alg:df2}
\KwData{$L$, $S$}
\KwResult{IsStateInteresting}

    \Comment{Interesting if the bucket changes from false to true}
    \If{$\exists {\rm\it Bucket} \in S$ s.t. ${\rm\it Bucket}$ changes from false to true} 
    {
    \Comment{Interesting if the location has been loaded before}
    \If{L at the index of ${\rm\it Bucket}$}{
        IsStateInteresting = true\;
    }
    }

\end{algorithm}


Since @load@ instructions which determine the interestingness of a memory location happen in the future, the determination of a @store@ instruction's ``interestingness'' at present is impossible. To resolve this issue, we propose to approximate the future interesting @load@ locations using past @load@ locations. The key insight here is that if a memory location has been written with interesting values in the past, then it is very likely to be written in the future. Therefore, \tool tracks the memory locations that were loaded in the past and the abstraction of the values being loaded.  If a store of a value in the current execution writes such a memory location and the abstraction of the value being written is different from all previous abstract values stored at the location, then we say that the store is interesting and the resulting state after the store is also interesting.

We show the dataflow waypoint algorithm in \Cref{alg:df1} and \Cref{alg:df2}.  During the execution of transactions, we maintain two maps (\Cref{alg:df1}) that record abstraction of past memory locations of @load@ and @store@ instructions separately. As discussed before, we use past load locations to determine whether a memory location is interesting. We can then check whether a store instruction in an execution is writing a new abstract value to the interesting memory location. Specifically, the check for whether a location is interesting or not happens as follows. Load map $L$ maps an abstract memory location to a Boolean value initialized to  @false@.  When a load instruction \verb|Load(Loc)| is executed, $L($\verb|Loc|$\ \%\ $\verb|MAP_SIZE|$)$ is set to true. @Loc@\ \%\ \verb|MAP_SIZE| is a course abstraction of @Loc@ which is necessary to create a small abstract address space. A small abstract address space helps lower storage and lookup overhead. Store map $S$ maps each abstract memory location to a \emph{bucket} abstraction of values. The bucket abstraction of a value is similar to the bucketing mechanism used in AFL \cite{afl}. Bucketing helps reduce the amount of total value inserted into the map, thus reducing the storage and evaluation cost, with the trade off for losing granularity of the inserted data. Each bucket is a partition of the domain of the actual value being stored, which is a 256-bit integer in the context of EVM. For instance, one partition plan for 256-bit integer could be $[2^0, 2^8), [2^8, 2^{16}), [2^24, 2^{256})$, which split the domain into three partitions. We use buckets to avoid storing the entire states, because storing all unique states is not effective. When a \verb|Store(Loc, Value)| is executed, \verb|Value| is first abstracted to determine its bucket. If $b \in B$ is the bucket abstraction of @Value@, $S($\verb|Loc|$\ \%\ $ \verb|MAP_SIZE|$)(b)$ is set to @true@. After execution, if a bucket in the store map $S$ changes from @false@ to @true@, and the corresponding location in $L($\verb|Loc|$\ \%\ $\verb|MAP_SIZE|$)$ is also @true@ (line 2 at \Cref{alg:df2}), then the new state is considered as interesting. The size and the range of each bucket can be fine-tuned as a hyper-parameter. With more number of buckets in each slot, more states would be evaluated to be interesting, increasing the size of $\mathcal{C}_s$. As size of $\mathcal{C}_s$ increases, it becomes harder to effectively choose the next state to explore and causes storage overhead. 



\subsection{Comparison Waypoint} \label{subsec:cmp}



        
  
\begin{algorithm}
\caption{Algorithm for comparison waypoint instrumentation}\label{alg:cmp1}
\KwResult{$C_{\text{local}}$}

$C_{\text{local}} \gets$ Vec<U256> = U256::max 

\While{EVM is Executing}{
  \If{Current Instruction is lt(Lhs, Rhs)}{
    $C_{\text{local}}$ (PC \% $\text{MAP\_SIZE}$)  $\gets$ Lhs - Rhs \;
  }
  \If{Current Instruction is eq(Lhs, Rhs)}{
    $C_{\text{local}}$ (PC \% $\text{MAP\_SIZE}$)  $\gets$ abs(Lhs - Rhs) \;
  }
  
}
\end{algorithm}

\begin{algorithm}
\caption{Algorithm for comparison waypoint evaluation}\label{alg:cmp2}
\KwData{$C$, $C_{\text{local}}$, $s$}

\For{$c\in C$, $c_{\text{local}} \in C_{\text{local}}$}{
  \Comment{When the current execution minimizes a slot in comparison map}
  \If{$c > c_{\text{local}}$}{
    $c \gets c_{\text{local}}$\;
    
    $s.votes = s.votes + 1$ \Comment{Vote for the state}
  }
}

\end{algorithm}
  
Since \tool is not sending a sequence of transactions to the smart contract, all intermediate states needed to construct a crucial state, must be included in either of the two corpuses. Consider the \verb|SimpleState| example in \Cref{fig:example}, if \tool only stores $0,1, 2$ in \verb|counter| and never saves $3$ for \verb|counter| because it falls in the range 2 to 4, which is already filled by 2, then \tool is never able to build up a state where \verb|counter| is 4 or even the bug-triggering state (where \verb|counter| is 10). If only dataflow waypoint is used, the inclusion of certain states may not be possible if the partition of the domain (i.e., buckets) is not fine-grained enough for the target smart contract.

Nevertheless, the problem with fine-grained partitioning is the significant amount of states in the infant state corpus, which leads to huge memory usage overhead over time for large smart contracts. To efficiently tackle this overhead, we propose to use \emph{comparison waypoints}. Comparison waypoints only considers all intermediate states, having some comparison instruction operands growing closer to each other than previous executions, which is required for reaching higher coverage, to be interesting. For instance, in \cref{fig:example}, comparison waypoint considers all states having @counter@ growing from 0 to 10 one by one as interesting because @counter@ grows towards the comparison target on line 16.

We show a simple algorithm in \Cref{alg:cmp1}  to build the comparison waypoint. First \tool initializes a map local to the current execution $C_{\text{local}}$ with maximum possible value (line 1). Then during execution, for every comparison instructions, \tool updates the distance at key $\text{Program Counter}\ \%\ \text{MAP\_SIZE}$ (i.e., the location of the comparison instruction) as in line 4 and 6. The distance reflects the proximity of two sides to achieve parity in a comparison, determined by the absolute value of their difference. For example, if \tool processes the \verb|EQ(1, 3)| operation, the distance would be 2. An execution is deemed interesting when it has a higher likelihood of any comparison being true than all previous ones. In another word, as depicted in \cref{alg:cmp2}, if the current execution displays a smaller distance in the local map ($C_{\text{local}}$) compared to the map that records the smallest distance ever recorded in previous executions ($C$), the execution is considered interesting.


Surprisingly, the comparison waypoint is useful to determine whether a state should be pruned from the infant state corpus to solve the memory overhead arising from fine-grained partitioning. Intuitively, to prune the infant state space using comparison waypoint, \tool simply prune states with less ``interestingness'' feedback from the comparison waypoints. In detail, we illustrate a voting algorithm in \cref{alg:cmp2} to track how interesting an infant is: each time when execution on an initial state (state before execution) minimizes the comparison map (line 2), \tool increases the votes for that initial state by 1. The number of votes encodes the interestingness of the state.

Now, recall the problem of the infant state corpus growing quickly when using dataflow waypoint with fine-grained partitions. This problem causes two issues:
\begin{enumerate}
\item There are too many states to select from for exploration and most states are likely similar to other states. A naive queue polling algorithm is not effective. (i.e., search space grows too large).
\item The infant state corpus grows too large to be stored in the fuzzing host. 
\end{enumerate}

To resolve issue (1), \tool needs to prioritize more interesting states in the infant corpus in terms of comparison waypoint. Naturally, we propose to use a probabilistic sampling algorithm that prioritizes the exploration of states that have greater votes. By intuition, if the future executions of the state minimize comparison map for more time, it should be selected more for exploration as it has greater potential to cover more code locations and generate other interesting states (according to the dataflow waypoint). To avoid being greedy (i.e., mostly exploring a specific state), the algorithm switches between random sampling and probabilistic sampling during each epoch. It is analogous to having two phases where one is in charge of maximizing rewards, and the other one is in charge of probing. 

To resolve issue (2), \tool prunes the infant state corpus when its size reaches a threshold. We want to prune least interesting yet most explored states inside the corpus. Similar to the previous solution, \tool sorts the states with respect to their \verb|votes/visits|, where @visits@ is the number of times this state has been chosen and executed. Then, \tool discards $M$ states with lowest \verb|votes/visits| and have \verb|visits| greater than a threshold $O$. If a state can not minimize comparison map over $O$ visits, then it is highly likely that either this state could be represented by another state or maximum coverage has been achieved for such a state regardless of how transactions are mutated. The pruning algorithm is expensive, but it is only called when the infant state corpus reaches the threshold, which only occurs less often during fuzzing.  


\cutreentrancy

\subsection{Reentrancy Exploit Synthesis}\label{subsec:reentrancy_syn}

During reentrancy attacks, external calls are leveraged to yield the control to attacker-controlled contracts, which can re-enter current or different functions for arbitrary times. The state during re-entry (i.e., partial state) includes all state changes conducted by instructions before the external call. Any instruction following the external call is only executed once re-entry calls finish. Thus, \tool can similarly store the snapshot of the state before external calls, like what \tool has done for states gained from fully executing a function. In this section, we first explain how storing partial executed states works, then we define the customized dataflow waypoint optimized for classifying a partial state.


To see an example, given a function $P$ containing a control leak, we can divide it into $P_{\text{pre}} \circ P_{\text{extcall}} \circ P_{\text{post}}$. Both $P_{\text{pre}}$ and $P_{\text{post}}$ are sequences of instructions and $P_{\text{extcall}}$ is the \verb|Call(Addr)| instruction causing the control leak. When $P_{\text{extcall}}$ is encountered during the execution and its target (\verb|Addr|) is unbounded, the fuzzer will snapshot the partial state up to $P_{\text{extcall}}$ and continue execution for the rest of the program by skiping $P_{\text{extcall}}$. After the execution, both the partial state $s'_{\text{partial}} = P_{\text{pre}} (s, t)$  and the finished state $s'_{\text{full}} = P_{\text{pre}} \circ P_{\text{post}} (s,t)$ will be evaluated by the waypoints and potentially be added to the infant state corpus. As the state derived from partial execution still requires further execution to complete the execution, \tool attaches the EVM stack and program counter to that state. 


In order to properly decide whether we should save a partial state into the corpus, we need to define fine-grained feedback for partial states as well. By intuition, a partial state is more likely to be useful when the instructions executed in $P_{\text{post}}$ modify the states that are checked by insctions executed before or during the external call. When it re-enters such a function or different functions, it may pass the checks again because instructions after the call that modifies the values in the check are not yet invoked. In order to find interesting partial states that lead to faster reentrancy exploit, we leverage such concept and define a variant of the dataflow waypoint (\Cref{sec:df}). More specifically, \tool tracks the dataflow waypoints for the full execution state ($s'_{\text{full}} = P_{\text{pre}} \circ P_{\text{post}} (s,t)$). A minor difference from the normal dataflow waypoint is that we separate the load/store map into $(L_{\text{pre}}, S_{\text{pre}})$ and $(L_{\text{post}}, S_{\text{post}})$ to track @load@ and @store@ instructions in $ P_{\text{pre}}$ and $P_{\text{post}}$, respectively. For the address set $l = \{{\rm\it addr}\ |\ L_{\text{pre}}({\rm\it addr}) = \texttt{true}\}$ and the address set $s = \{{\rm\it addr}\ |\ S_{\text{post}}({\rm\it addr}) = \texttt{true}\}$, if there exists an ${\rm\it addr}$ in $s$ such that $ {\rm\it addr}$ is also in $l$, we consider the state derived from partial execution to be interesting. Why this waypoint leads to efficient reentrancy exploit synthesis? Intuitively, reentrancy at a specific $P_{\text{extcall}}$ is highly exploitable when the $P_{\text{post}}$ modifies part of the state read by $P_{\text{pre}}$. Otherwise, re-entering the current function $P_{\text{pre}}$ would not have any effect in comparison to simply re-execute the function multiple times, which is already handled by the fuzzer. The newly added reentrancy-specific waypoint is orthogonal to all previous waypoints, i.e., we track the uniqueness of difference between $L_{\text{pre}}$ and $S_{\text{post}}$ and unique dataflow pattern globally, which is conducted by the dataflow waypoint described in \Cref{sec:df}.

\fi

\section{Implementation} \label{sec:impl}

To implement the snapshot-based fuzzing algorithm and both the dataflow and comparison waypoints, we build \tool from scratch. We use LibAFL \cite{libafl} as a backbone and implement a separate state corpus to support snapshotting the states. We also incorporate the dataflow and comparison waypoints into \tool using customized feedback. Because \tool is implemented in Rust, we use revm \cite{revm} as the EVM executor for convenience. We also leverage revm's interpreter hook to perform dynamic instrumentation, collect dataflow and comparison information, and conduct fast snapshotting. \tool also supports starting from a state pulled from a specific block from any blockchains supporting Geth client. 

\tool is modularized. It can be easily extended to support fuzzing smart contracts on other chains like Solana by adding new executor components. New domain-specific feedback or waypoints can also be added quite easily. \tool is opensourced at \url{https://github.com/fuzzland/ityfuzz}

\section{Evaluation} \label{sec:eval}

\paragraph{Research questions} 
In this section, we show the performance brought by snapshot-based fuzzing and various waypoint mechanisms. We aim to answer the following research questions:
\begin{enumerate}[label=\textbf{RQ\,\arabic*:}, ref={RQ\,\arabic*}]
\item Does \tool perform better than other tools with regards to instruction coverage? \label{RQ1}
\item Can \tool identify and generate exploit for real-world vulnerabilities? \label{RQ2}
\item Does storing state instead of a sequence of transactions incur high memory overhead and can this be resolved by using waypoints? \label{RQ3}
\item Is on-chain auditing beneficial for uncovering vulnerabilities ignored during the development cycle? \label{RQ4}
\item Can the response time of \tool support on-chain auditing? \label{RQ5}
\end{enumerate}

\paragraph{Experimental setup}

We leverage three datasets (B1, B2, and B3) to evaluate our tool. B1 (extracted from VERISMART \cite{DBLP:conf/sp/SoLPLO20}) contains 57 smart contracts supporting ERC20 standard interface (i.e., tokens). B2 and B3 contains 72 and 500 smart contracts crawled from the Ethereum chain. We build \tool as described in \Cref{sec:impl}, and to evaluate the effectiveness of each technique, we create two ablations of \tool:
\begin{itemize}[leftmargin=1.5em]
\item \tool - Rand: snapshots states and stores them into the infant state corpus with a likelihood of 50\%.
\item \tool - DF: snapshots states and stores them into the infant state corpus based on only dataflow waypoint.

\end{itemize}

We perform all experiments on a machine with AMD Epyc CPUs (128 cores) and 256 GB memory. All ablations and other tools used in the evaluation are compiled with optimization (-O3).

\subsection{Coverage}
To answer \ref{RQ1}, whether our tool performs well on coverage metrics, we compare our tool with the state-of-the-art smart contract fuzzer SMARTIAN \cite{DBLP:conf/kbse/0001K0GGC21} using dataset B1, B2, and B3. We do not compare with other fuzzers mentioned in \Cref{sec:related} because SMARTIAN has already been shown to significantly outperform all these tools. For each dataset, only instruction coverage is used for comparison since control flow cannot represent of statefulness of smart contract programs. We count the total possible coverage by summing number of instructions in all deployed contract bytecode. As neither our tool nor SMARTIAN supports mutating function hash and generating spurious input data not conforming to the structural requirement, we do not count instructions in the basic block that validates input structure and function hash. We also removed unreachable code (i.e., metadata) that present after the last \verb|Return| instruction in the program.

\begin{figure}[ht]
\begin{minipage}[t]{.47\textwidth}
        \centering
        \includegraphics[width=\textwidth]{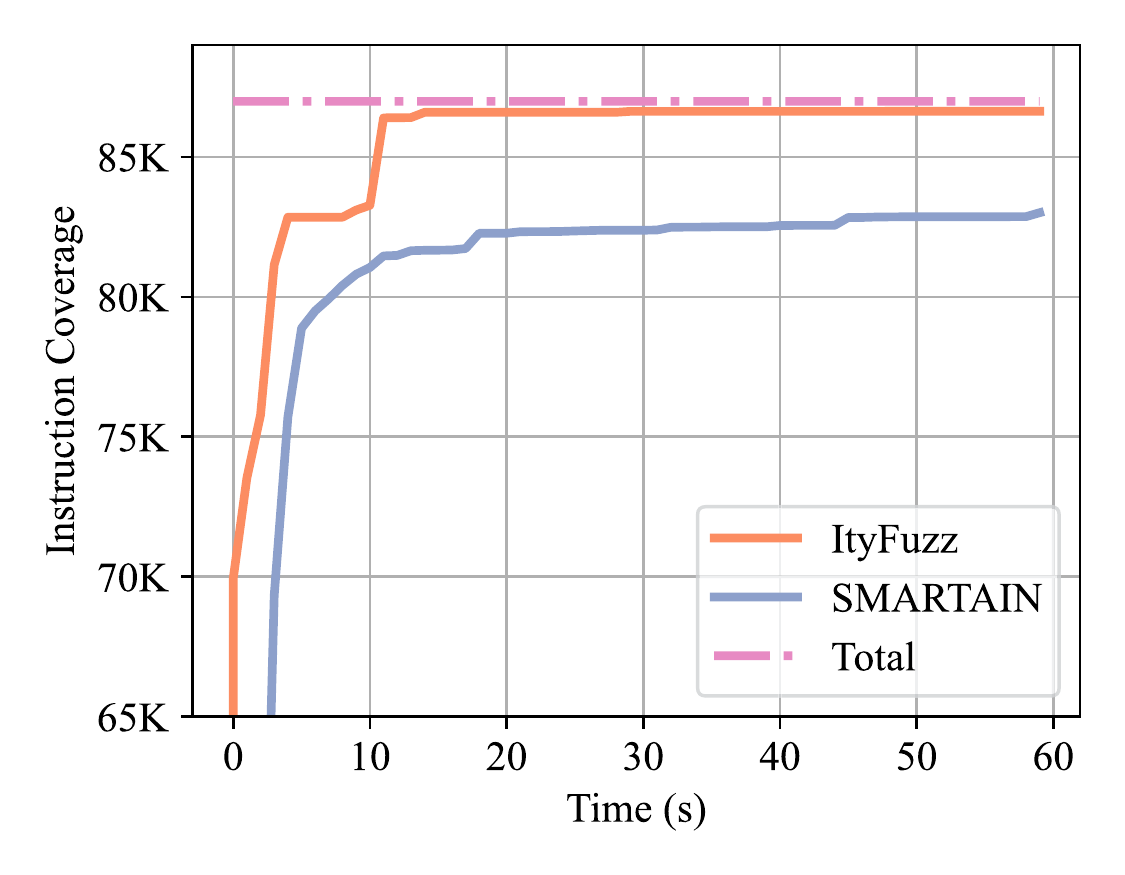}
        \caption{Instruction coverage for different tools over time for B1, Total is the total number of instructions, higher is better}
        \label{fig:icb1}
\end{minipage}\hfill
\end{figure}

\begin{figure}[ht]
\begin{minipage}[t]{.47\textwidth}
        \centering
        \includegraphics[width=\textwidth]{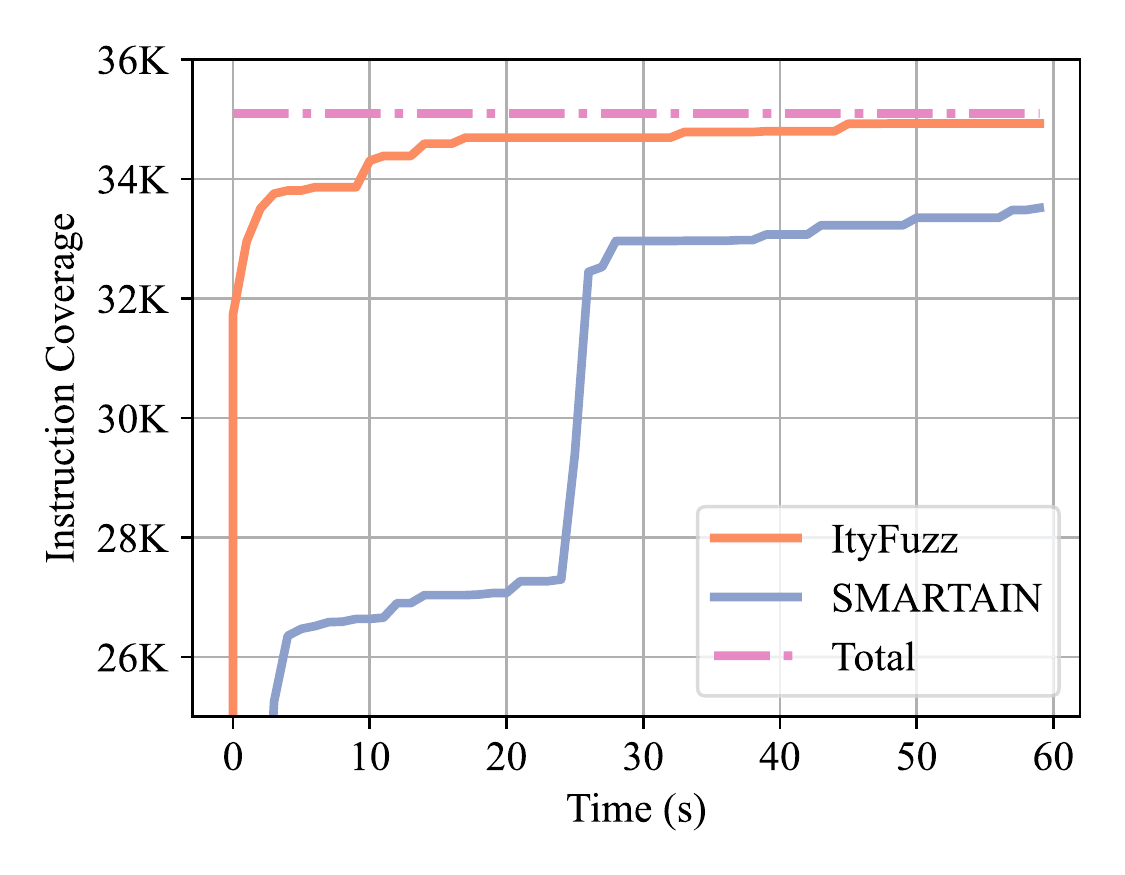}
        \caption{Instruction coverage for different tools over time for B2, Total is the total number of instructions, higher is better}
        \label{fig:icb2}
\end{minipage}
\end{figure}

\begin{figure}[ht]
\begin{minipage}[t]{.47\textwidth}
        \centering
        \includegraphics[width=\textwidth]{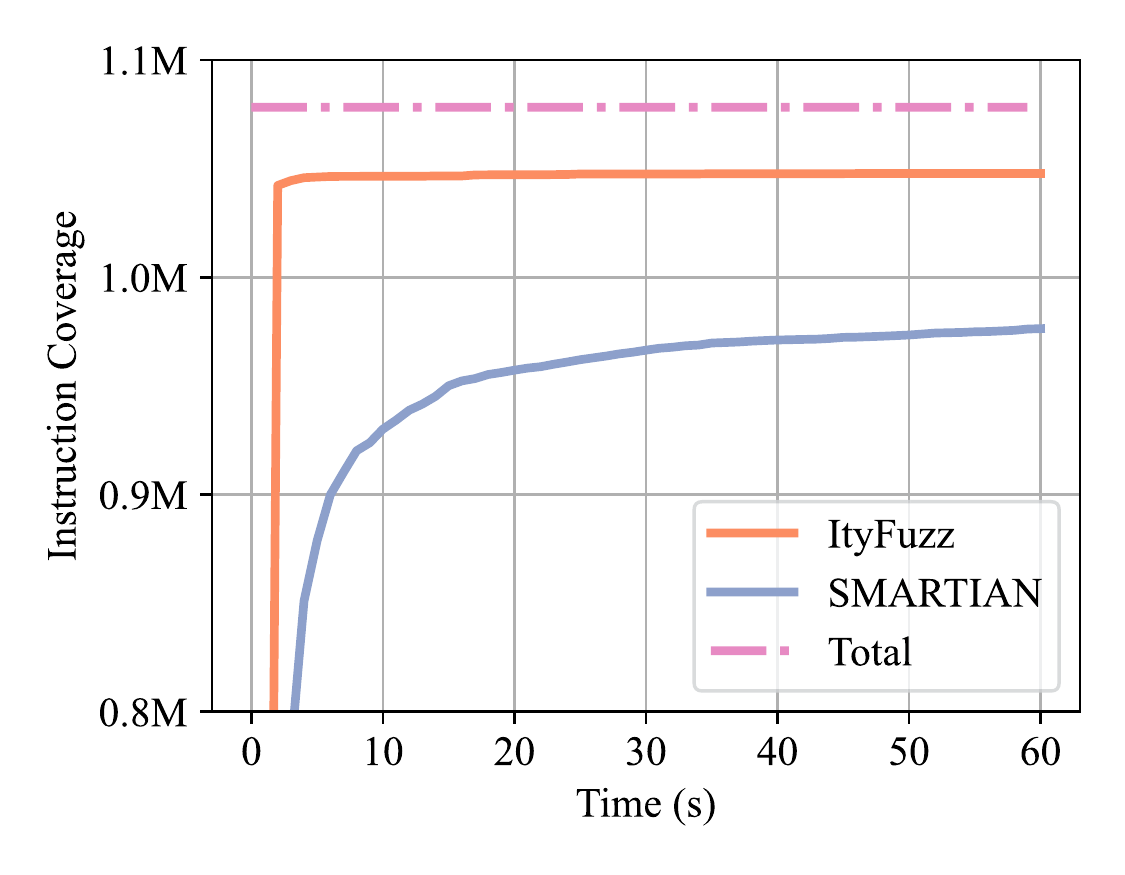}
        \caption{Instruction coverage for different tools over time for B3, Total is the total number of instructions, higher is better}
        \label{fig:icb3}
\end{minipage}
\end{figure}

\Cref{fig:icb1}, \Cref{fig:icb2}, \Cref{fig:icb3} show the total instruction coverage over time for each tool on the B1, B2, and B3 datasets respectively. Even though SMARTIAN uses concolic execution, \tool performs significantly better than SMARTIAN regarding the instruction coverage versus time. Time improvement can be justified by using snapshots of states instead of building up states with re-execution, as re-execution incurs significant time overhead. For the smart contracts in B1, \tool covers almost all instructions after 10 seconds while SMARTIAN can not do so in one minute. This is because comparison waypoints prioritize explorations of states that can potentially improve future coverage.


\subsection{Vulnerabilities}


To  answer \ref{RQ2}, we tested \tool on real-world smart contract projects. We gathered 42 previously exploited projects and fuzzed each project with a one-hour time limit. Among them, \tool was able to identify concrete exploits for 36 of them, with an average time of 13.8 seconds. SMARTIAN failed to reach vulnerable states in 24 hours for Bacon Protocol and EGD Finance (determined by our custom oracle). Due to lack of automation, creating oracles becomes tedious and almost impossible for all 42 projects. Additionally, reputable firms audited most of the 42 projects, further showcasing our tool's effectiveness. There is no false positive as \tool identified actual exploits that could be executed on a fork. The main reason for false negatives was that certain projects required input of an address (160 bits) or a signature (>256 bits) with no hints (i.e., no comparison with a constant), making it difficult to brute-force using the fuzzer. We believe that concolic execution can resolve this problem easily and leave it as potential future work. 

We also applied our tool to 45000 smart contract projects (with more than 150k smart contracts) received >100 transactions on both Binance Smart Chain and Ethereum. Fuzzing each project for five minutes, ItyFuzz is able to generate concrete exploits for stealing assets valued at over \$500k (approximated using Uniswap data) among 21 vulnerable projects, consisting of liquidity pools, ERC20 tokens, and a GameFi. Additionally, ItyFuzz can also find common vulnerabilities like arbitrary external calls and arbitrary ERC20 token burning in 1384 projects, holding assets valued at over \$8M. 

We are unable to compare our tools with previous works. All existing tools do not support multiple-contract fuzzing, require huge amount of manual effort to create harness and invariant, or can not generate concrete exploits (e.g., only report all potential bug locations via static analysis). Thus, we present an ablation study with each project and report the time to identify the bug. The result is shown in \Cref{table:abvuln}.

\begin{table*}[]
\caption{Ablation Study of Vulnerability Detection Time, OOM stands for out of memory}

\begin{tabular}{|l|l|l|l|l|l|}
\hline
\textbf{Project} & \textbf{Exploit Type} & \textbf{\tool Rand} & \textbf{\tool DF} & \textbf{\tool}  \\ \hline
DVD - Unstoppable & DoS & 1.4s & Timeout &  1.3s \\ \hline
Bacon Protocol & Reentrancy & OOM & Timeout &16.4s \\ \hline
N00d Token & Reentrancy & OOM & Timeout & 28.2s \\ \hline
EGD Finance & Price Manipulation & OOM & 6.2min & 9.3s \\ \hline
Contract 1 (Undisclosed) & Access Control & 0.7s & 0.6s & 0.7s \\ \hline
Contract 2 (Undisclosed) & Reentrancy & OOM & 54.7s & 1.6s \\ \hline

\end{tabular}%
\label{table:abvuln}
\end{table*}

For most vulnerabilities, \tool outperforms all its ablations. Especially when the contract is complex and has deep state space, \tool can find the vulnerabilities in a short time while \tool - Rand and \tool - DF time out after one hour.

Rarely, \tool - DF times out while \tool - Rand does not. This is because \tool - DF, based on dataflow waypoints (\Cref{sec:df}) may fail to include a crucial state to build up the final state of which the vulnerability is found. This happens because a state is only added to the corpus when its dataflow value falls into a unique bucket. States falling into the same bucket, despite with different values, are ignored although sometimes such states are interesting and important. On other hand, \tool Rand does not block any state from inclusion into the corpus. \tool prevents this issue by providing higher granularity for what states are considered to be interesting, allowing these states ignored by \tool DF to be stored in the infant state corpus, thus facilitating the exploration.


\subsection{State Overhead}\label{subsec:stateoverhead}

\begin{figure}[ht]
        \centering
        \includegraphics[width=230pt]{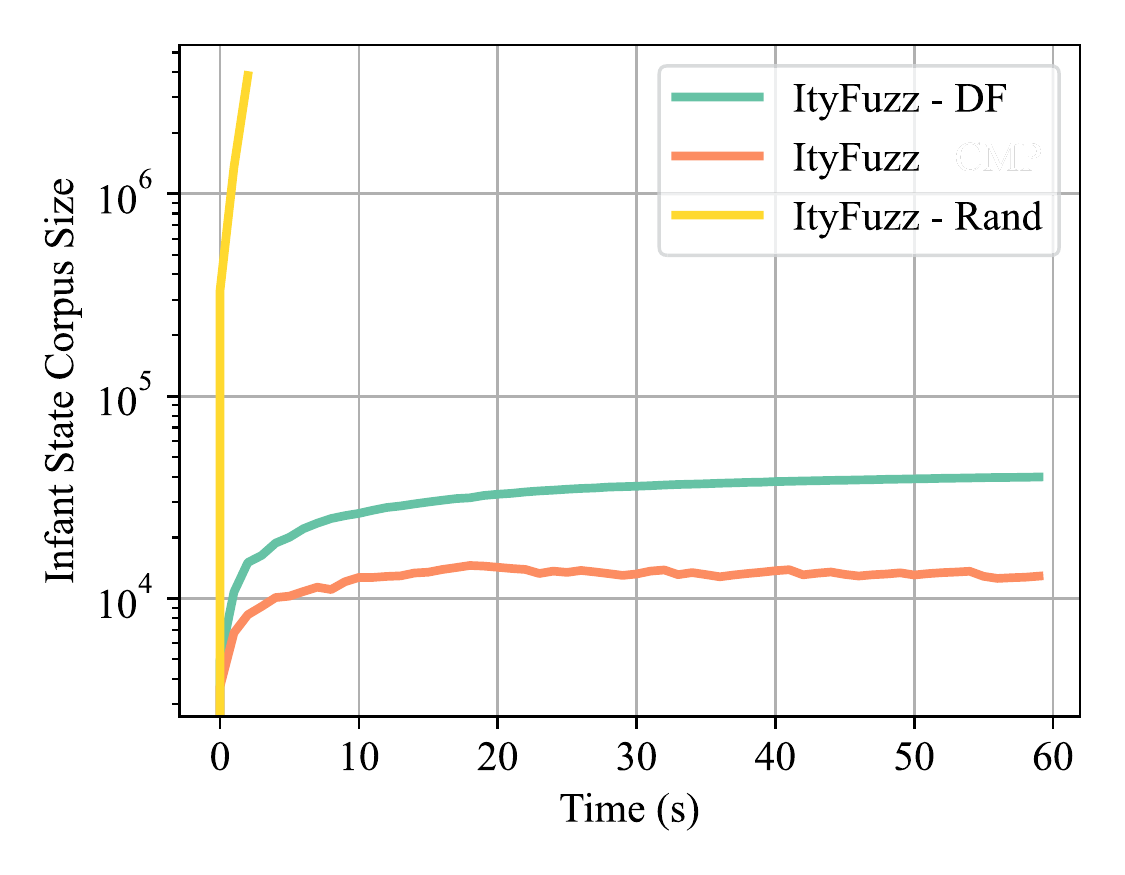}
        \caption{Infant State Corpus Storage Overhead, Y axis is the number of unique states stored in the infant state corpus}
        \label{fig:icb0}
\end{figure}

To answer \ref{RQ3}, whether storing states incur memory overhead, we conduct an ablation study of \tool. We count the total amount of items in the infant state corpus over time for executing \tool and its ablations on dataset B1. The result is shown in \Cref{fig:icb0}.

\tool - Rand crashes because of out-of-memory (OOM) after approximately three seconds for most smart contracts. This happens because \tool - Rand keeps adding states into the infant state corpus, meaning that a constant fraction of all execution resultant states are stored in the corpus. On the other hand, states added to the corpus of \tool - DF increases gradually for the first 20 seconds and converges to an approximate constant. This is because at first, the dataflow waypoint collects the states that store to interesting locations according to the load map and set the corresponding entry in the store map's bucket to true. After that, states are less likely to be added to the corpus unless the load map changes. This is because if the state does store at an interesting location, but the store map entry is already set to true, we do not save the state to corpus again. \tool actively prunes the infant state corpus so that the size ultimately converges to a constant.

\subsection{On-Chain Auditing} \label{subsec:onchain}

\begin{table*}[]
\caption{Vulnerability Detection Time}
\begin{tabular}{|l|l|l|l|l|}
\hline
\textbf{Project} &\textbf{ Exploit Type} & \textbf{Reaction Time} &\textbf{ \tool (Dev)} & \textbf{\tool (On-chain)}  \\ \hline
Nomad Bridge & Incorrect Initialization & 41 Days & Timeout & 0.3s \\ \hline
Team Finance & Logic Flaw & 1.12 Hour & Timeout & 2.2s  \\ \hline
\end{tabular}%

\label{table:fastvuln2}

\end{table*}

To answer the last two research questions \ref{RQ4} and \ref{RQ5}, whether fast vulnerability discovery can prevent the attack, we evaluate our implementation on two previously hacked DeFi projects: Nomad Bridge and Team Finance. The result is shown in \Cref{table:fastvuln2}.\newline

\textit{Nomad Bridge} is a cross-chain bridge allowing for fund transfer. On August 1st, 2022, 41 days after the vulnerable contract is deployed, attackers exploited the bridge and stole \$190M assets. The vulnerability occurred because of an incorrect initialization on the chain. 

When we ran \tool on their test deployment script from commit hash \verb|3a997a44| (i.e., one day before the attack happens), \tool could not find the vulnerability. Similarly, CI for \verb|3a997a44|, which contains dynamic and static analysis, had not reported any vulnerability. This is because their deployment script initializes correctly. If we fork the actual chain from block 15259103 (after the deployment of the new contracts) and perform fuzzing based on that state, \tool can identify the vulnerability in 0.3s. This example illustrates that testing in the development environment is not enough, especially for code that manages huge amounts of real-world assets. As the production environment may differ from the testing and staging environment, using \tool - On-chain can account for these differences. 

Moreover, as the techniques adopted by hackers are also evolving, such a vulnerability is likely to be discovered and exploited in a few minutes. As we reduced the vulnerability discovery time to a sufficiently low value (0.3s), contract deployers can leverage our tool to continuously monitor after the deployment, and gain concrete exploitation information before the hackers do. \newline

\textit{Team Finance} is a DeFi platform for token storing and vesting. It was hacked on October 7th, 2022 and hackers stole \$15.8M worth of assets \cite{TeamFinance}. The vulnerability is simple yet ignored by manual auditors. Specifically, in their migrate fund contract, there is a vulnerable function \verb|migrate| (\Cref{fig:teamfinance}), whose argument \verb|sqrtPriceX96| controls the swapping rate of the liquidity pool (i.e., the argument can control the swapping price between two tokens). \verb|sqrtPriceX96| should only be gained from trusted entities, not arbitrary callers. If a hacker creates a skewed swapping rate via \verb|sqrtPriceX96| between a token they can control and a token with real value, then they can generate huge refunds of that token with real value from the migration and earn from the refunds. 

\tool cannot find it offline in one hour because building up a state that allows conditions in lines 6 to 9 to be true is non-trivial. To build such a state, the message sender has first to gain balance $k$ of tokens, approve contract with $k'$ token, deposit into the contract through calling \verb|lockToken| with the correct amount of ETH sent for locking fee and $v$, such that $v < k' \wedge v < k$, and finally call \verb|extendLockDuration| with the correct ID gained from \verb|lockToken| and the timestamp to increase locking time greater than current time stamp. After reaching this state, hackers also need to generate a correct \verb|sqrtPriceX96| to profit from the attack. Still, given a longer time to run and build up states, \tool can discover such a vulnerability. However, longer vulnerability discovery time is not desirable for contract deployers.

In reality, the attack initiated by a hacker at block 15837893 (Oct-27-2022 07:22:59 AM +UTC) (creating a token, transferring balance, and locking the amount) finalized at block 15838225 (Oct-27-2022 08:29:23 AM +UTC). If we start from block 15837893 and use the attacker address as the address controlled by the \tool, \tool can identify the vulnerability in 2.2s on average, which is less than the time (1 hour 7 minutes) before the exploit finalizes. \tool - On-chain can give developers enough time to prevent the final harmful transaction from being successfully executed by pausing the contract or even fixing the vulnerability. As a result, \tool - On-chain is beneficial for exploring previously unexplored code locations made possible by the attackers. 

\begin{figure}
\begin{lstlisting}[language=Solidity, xleftmargin=3.5em, xrightmargin=0.5em]
function migrate(uint256 _id, IV3Migrator.MigrateParams calldata params,
    bool noLiquidity, uint160 sqrtPriceX96, bool _mintNFT
) {
    ...
    Items memory lockedERC20 = lockedToken[_id];
    require(block.timestamp < lockedERC20.unlockTime, "Unlock time already reached");
    require(_msgSender() == lockedERC20.withdrawalAddress, "Unauthorised sender");
    ...
    uint256 token0BalanceBefore = IERC20(params.token0).balanceOf(address(this));
    uint256 token1BalanceBefore = IERC20(params.token1).balanceOf(address(this));
    //initialize the pool if not yet initialized
    if(noLiquidity) {
        v3Migrator.createAndInitializePoolIfNecessary(params.token0, params.token1, params.fee, sqrtPriceX96);
    }
    v3Migrator.migrate(params);
    //refund eth or tokens
    ...
}
\end{lstlisting}
    \caption{Migrate function of Team Finance}
    \label{fig:teamfinance}
\end{figure}

Lastly, there has not yet been any attack that manipulates price oracle on EVM-based chains. However, Mango Market, which is a Solana smart contract project, has been recently exploited because the attacker controlled a price oracle, and the smart contract become vulnerable after the price is skewed. \tool - On-chain can prevent these vulnerabilities since it is based on a real-time price oracle and can conduct analysis when the price becomes skewed. We envision the number of vulnerabilities related to oracle manipulation will grow over time as there are more and more state-sponsored hackers, who are equipped with assets enough to manipulate oracles, joined such a realm \cite{nk_sanction} \cite{lutz_north_2022}.

\section{Related Work}\label{sec:related}
\subsection{Feedback-driven Fuzzing}
\paragraph{Coverage-guided fuzzing}

To increase the search efficiency for fuzzers, various feedbacks are collected dynamically during the execution process. One of the most popular and effective feedback is coverage, i.e., how many instructions or control flow edges have been covered by the current execution. By prioritizing inputs that led to new coverage, the fuzzers can widely explore the whole execution space. Most, if not all, modern software fuzzers (e.g. AFL \cite{afl}, HonggFuzz \cite{HonggFuzz}, FairFuzz \cite{DBLP:conf/kbse/LemieuxS18}) incorporate coverage-feedback into the search process. However, coverage feedback is not always effective for stateful smart contracts. For example, in the \Cref{fig:example} example, almost all smart contract fuzzers can easily reach all coverage in both @incr@ and @decr@ functions, but since our goal (bug) is embedded in the @buggy@ function with a hard-to-solve stateful guard, coverage-guided fuzzers are not able to proceed further.

\paragraph{Customized Waypoints}

To address the coverage-guided fuzzing issue, some fuzzers introduce more exotic feedback. For example, Validity fuzzing \cite{DBLP:conf/icse/PadhyeLSPT19} incorporates the feedback of the validity of the inputs with respect to the target program, e.g. whether the input has a valid HTTP header to a web server. SlowFuzz \cite{DBLP:conf/ccs/PetsiosZKJ17} tracks the execution path length in order to prioritize inputs that lead to deeper execution spaces. To find performance bugs, PerfFuzz \cite{DBLP:conf/issta/LemieuxPSS18} improves SlowFuzz by taking multi-dimensional feedback from the execution including execution counts at all program locations. Another powerful feedback extensively used by the security community is the comparison feedback \cite{DBLP:conf/sigsoft/LiCCLLT17, 8418632,Rawat2017VUzzerAE,DBLP:conf/uss/Yun0XJK18}. In real-world software, many hard comparison constraints are difficult to solve, and comparison feedback partially solves this issue by recording inputs that are closer to the hard condition constants. Finally, FuzzFatory\cite{DBLP:journals/pacmpl/PadhyeLSSV19} formalizes the concept of waypoints which provide customization of feedbacks. However, all these feedbacks and waypoints are designed for general traditional software which differs substantially from smart contracts.
Inspired by the above work, we propose two novel feedback mechanisms tailored for stateful smart contract programs in \Cref{sec:df} and \Cref{subsec:cmp} and demonstrate their effectiveness in \Cref{sec:eval}.

\subsection{Stateful Fuzzing}
SMARTIAN \cite{DBLP:conf/kbse/0001K0GGC21} starts from a clean state and sends a sequence of transactions to build up the state. Nyx \cite{nyx} implements a fast OS-level snapshotting strategy and subsequent work Nyx-Net \cite{nyxnet} leverages this snapshotting techniques to reset the state for complex stateful targets efficiently. Additionally Nyx-Net stores snapshots incrementally (i.e., stores snapshots after executing every K inputs). Similarly, Dong et al. incrementally snapshots Android OS for effective time travelling to previous state during testing Android applications. Storing snapshots every K inputs is impractical in our scenarios, as illustrated by \cref{subsec:stateoverhead}. To resolve state explosion, \tool stores states in infant state corpus only when waypoints consider the state as interesting. \tool also leverages schedulers for polling the corpus and corpus culling techniques to further ensure we are exploring interesting states. CorbFuzz \cite{corbfuzz} is a web application fuzzer. It tackles the statefulness of web applications by modeling the states and synthesizing a result for each data load query. These synthesized data can be spurious since they are not built from concrete store queries. Yet, \tool builds the state from concrete transactions and each state can be reproduced via a sequence of transactions. It is impossible for \tool to generate false positives. 

\subsection{Smart Contract Security Tools}
To secure economically crucial smart contract applications, various fuzzers and dynamic analysis tools have been developed to detect security bugs in smart contract programs. ContractFuzzer \cite{DBLP:conf/kbse/0001LC18} is a black-box contract fuzzer with relatively low overall coverage. Echidna \cite{DBLP:conf/issta/GriecoSCFG20} and Harvey \cite{DBLP:conf/sigsoft/WustholzC20} are two industrial-adopted fuzzers. More recently, SMARTIAN \cite{DBLP:conf/kbse/0001K0GGC21} is a hybrid smart contract fuzzer that also leverages static and dynamic data-flow analysis. However, all these tools fail to effectively reuse intermediate states and thus have large re-execution overhead. We solve this issue by introducing a novel snapshot-based fuzzing technique (\Cref{subsec:snapshot}) and develop a blazing fast fuzzer called \tool.

Beyond fuzzers, other program analysis tools have been developed for smart contracts as well, including symbolic execution tools like Mythril \cite{Mythril2018}, Manticore \cite{DBLP:conf/kbse/MossbergMHGGFBD19}, verification approaches \cite{DBLP:conf/cpp/AmaniBBS18, DBLP:conf/vstte/HajduJ19, DBLP:conf/ndss/KalraGDS18}, static analysis frameworks \cite{DBLP:journals/corr/abs-1809-03981, DBLP:conf/esop/ChatterjeeGV18,  DBLP:journals/pacmpl/GrechKJBSS18, DBLP:conf/icse/FeistGG19}. A common problem among these tools is the human-in-the-loop requirement, which is less desired because real-world smart contract exploits are time-sensitive and only highly efficient automated tools can detect them in time. On the other hand, \tool is a fully automated fuzzer that can detect security bugs in smart contracts without human intervention in only a few seconds.





\section{Conclusion}
In conclusion, we design a new snapshot-based fuzzer \tool for testing smart contracts that effectively stores intermediate states to reduce re-execution overhead. We define multiple customized waypoint mechanisms to efficient categorize and store interesting states for better program explorations. We also demonstrate state snapshots enable fast reentrancy exploits synthesis. Finally, we show that with our low response time fuzzer \tool, we can perform on-chain auditing to identity and prevent exploits for real-world smart contract applications.

\section*{Acknowledgements}
This work was supported in part by NSF grants CCF-1900968, CCF-1908870, and CNS1817122 and SKY Lab industrial sponsors and affiliates Astronomer, Google, IBM, Intel, Lacework, Microsoft, Mohamed Bin Zayed University of Artificial Intelligence, Nexla, Samsung SDS, Uber, and VMware. Any opinions, findings, conclusions, or recommendations in this paper are solely those of the authors and do not necessarily reflect the position or the policy of the sponsors.

\bibliographystyle{ACM-Reference-Format}
\bibliography{reference}
 
\end{document}